\documentclass[12pt]{article}
\usepackage{amsmath,amsthm}
\usepackage{eufrak}
\usepackage[pdftex]{graphicx}

\DeclareMathAlphabet{\bb}{U}{msb}{m}{n} \gdef\C{\bb C}    \gdef\dS{\bb S} \gdef\R{\bb R}
 \gdef\BH{\bb H}  

 \DeclareMathOperator{\spin}{{\bf
Spin}}

 \DeclareMathOperator{\Mat}{Mat}

\DeclareMathOperator{\SO}{SO}

\newcommand{\scr}{\scriptstyle}

\newcommand{\cA}{\mathcal{A}}

\newcommand{\bx}{{\bf x}}

\newcommand{\bZ}{{\bf Z}}

\newcommand{\Lip}{\boldsymbol{\Gamma}}

\newcommand{\cl}{C\kern -0.2em \ell}

\newcommand{\e}{\mbox{\bf e}}

\begin{document}
\title{Rozenfeld's Geometric Approach to Spinors}
\author{V.~V. Varlamov\thanks{Siberian State Industrial University,
Kirova 42, Novokuznetsk 654007, Russia, e-mail:
varlamov@sibsiu.ru}}
\date{}
\maketitle
\begin{abstract}
Rosenfeld's geometric approach to spinors is considered, according to which the coordinates of spinors are represented by the coordinates of the plane generators of the maximal dimension of the absolutes of non-Euclidean spaces. As an example, non-Euclidean spaces with neutral signatures are investigated.
\end{abstract}
{\bf Keywords}: spinors, Clifford algebras, non-Euclidean spaces, absolute, plane generators
\date{}
\maketitle

\section{Introduction}
Michael Atiyah's statement about spinors is widely known: "No one fully understand spinors. Their algebra is formally understood but their geometrical significance is mysterious" \cite[P.~430]{FG}. However, in the mid-50s of the last century, B.A. Rosenfeld proposed a very interesting geometric approach to the description of spinors \cite{Rozen55}. This approach provides a very clear and consistent geometric interpretation of spinors. Unfortunately, the book \cite{Rozen55} was never translated into English, and Rosenfeld's approach turned out to be virtually unknown to the English-speaking audience. This article aims to fill this gap.

As Atiyah said, the algebra of spinors is formally clear to us. The language of this algebra is based on the theory of Clifford algebras. The roots of Clifford algebra theory go back to Hamilton \textit{quaternions} \cite{Ham} and Grassmann {\it Ausdehnungslehre} \cite{Grass}. The algebra introduced by Clifford \cite{Cliff} is a generalization of the quaternion algebra to the case of multidimensional spaces. As a consequence of this generalization, a quaternion structure arises in the Clifford algebra, which is a tensor product of quaternion algebras, that is, a tensor product of four-dimensional algebras (the dimension of the quaternion algebra is 4). Studying the rotations of the $n$-dimensional Euclidean space $\R^n$, Lipschitz \cite{Lips} found that the group of rotations of the space $\R^n$ with the determinant $+1$ is represented by a \textit{spinor group}. As is known \cite{Rozen55}, the groups of motions of $n$-dimensional non-Euclidean spaces $S^n$ are isomorphic to the groups of rotations of spaces $\R^{n+1}$. Since Clifford algebras are isomorphic to matrix algebras, \textit{spinor representations} of motions of spaces $S^n$ can be considered as representations of vectors in corresponding spaces. The vectors of these spaces are called \textit{spinors} of spaces $S^n$. The concept of spinor was introduced by Cartan \cite{Car13}. Van der Waerden notes \cite{Waer} that the name ``spinor'' was given by Ehrenfest with the appearance of the famous article of Uhlenbeck and Goudsmit \cite{UG} about a spinning electron. More precisely, the geometric meaning of the spinor representations of the motions of non-Euclidean spaces $S^n$ is that the coordinates of the spinors can be considered as the coordinates of the plane generators of the maximal dimension of the absolutes\footnote{An \textit{absolute} is a set of infinitely distant points of a non-Euclidean space.} of these spaces, and the spinor representations of the motion of these spaces coincide with those transformations of the spinors that correspond to the transformations of the absolutes during the movements (see section 4). Thus, in a case important for physics, it is known that the connected group of motions of the three-dimensional non-Euclidean space $S^{1,2}$ (Lobachevsky space) is isomorphic to the connected group of rotations of the four-dimensional pseudo-Euclidean space $\R^{1,3}$ (Minkowski space-time), coinciding with the group of Lorentz transformations of special relativity. Therefore, the spinor representations of the connected group of motions of the space $S^{1,2}$ are at the same time spinor representations of the Lorentz group. It follows that each spinor of the space $S^{1,2}$ corresponds to some point of the absolute\footnote{The absolute of the Lobachevsky space $S^{1,2}$ is homeomorphic to the extended complex plane $\C\cup\{\infty\}$.} of the space $S^{1,2}$, and each point of the absolute of the space $S^{1,2}$ corresponds to an isotropic line of the space $\R^{1,3}$ passing through some point of this space. The described geometric interpretation of spinors and spinor representations was proposed by Cartan \cite{Car38} (see also \cite{Rozen55}).

In his book \cite{Che54}, Chevalley notes that the construction of the concept of spinor given by Cartan was rather complicated, and a much simpler presentation of the theory, based on the use of Clifford algebras, was given by Brauer and Weyl in \cite{BW35}. In his presentation of the algebraic theory of spinors, Chevalley follows the article \cite{BW35}. Thus, along with the geometrical interpretation, an algebraic approach to the description of spinors and spinor representations appeared, which was further developed in \cite{Port,Cru,Lou}. The productivity and development of the algebraic approach shows that the spinor is primarily an object of algebraic nature. According to the algebraic definition, a spinor is an element of the minimal left ideal of the Clifford algebra\footnote{The first one to consider spinors as elements in a minimal left ideal of a Clifford algebra was M. Riesz \cite{Rie47}.} $\cl(V,Q)$, where $V$ is a vector space equipped with a non-degenerate quadratic form $Q$. For $n$ even, the minimal left ideal of $\cl(V,Q)$ corresponds to the \textit{maximal totally isotropic subspace}\footnote{A subspace $U$ of a space $V$ is called totally isotropic if the bilinear form $B(\alpha_i,\alpha_j)=0$ for all $\alpha_i,\alpha_j\in U$. A subspace $U\subset V$ of maximal dimension with the above property is called a maximal totally isotropic subspace.} $U\subset V$ of dimension $n/2$, i.e. it is isomorphic to a spinspace $\dS$ of dimension $2^{n/2}$ \cite{Abl01}.

\section{The Absolute and Ideal Points\\ of non-Euclidean Spaces}
According to Rozenfeld \cite{Rozen55}, non-Euclidean spaces $S_{p,q}$ are defined as hyperspheres of real pseudo-Euclidean spaces $\R_{p+1,q}$ with diametrically opposed points identified. Each such pair of points corresponds uniquely to a straight line of space $\R_{p+1,q}$ passing through the center hypersphere and connecting these points. Therefore, each point of the space $S_{p,q}$ corresponds to some straight line of the bundle of lines of the space $\R_{p+1,q}$. In this case, \textit{the distance between the points of non-Euclidean space is equal to the absolute value of the product of the angle between the corresponding lines by a constant real or purely imaginary number -- the radius of curvature of space}. Thus, we obtain interpretations of non-Euclidean spaces in bundles of lines of spaces $\R_{p+1,q}$.

Since not every straight bundle of lines passing through the center of the hypersphere of the space $\R_{p+1,q}$ intersects with this hypersphere and pairs of diametrically opposed points of the hypersphere of real or imaginary radius are in one-to-one correspondence only with a part of the straight bundles, \textit{spaces $S_{p,q}$ of negative or positive curvature are represented by those straight bundles that intersect with the hypersphere of the real or purely imaginary radius}.

Interpreations of non-Euclidean spaces in bundles of straight lines allow us to expand these spaces: we will consider straight bundles that do not intersect with the hypersphere to represent some new point of this space. We will call these new points corresponding to straight lines crossing the hypersphere at imaginary points \textit{ideal points} of non-Euclidean space, and the totality of such points is \textit{an ideal domain} of this space. The points of non-Euclidean spaces that we dealt earlier will be called \textit{proper points} of non-Euclidean space, and the totality of such points will be called \textit{the proper domain} of this space. The set of points bordering between these two types of points corresponding to the asymptotes of the hypersphere will be called \textit{the absolute} of non-Euclidean space. We will call a non-Euclidean space, together with its absolute and ideal domain, \textit{an extended non-Euclidean space}. The distance between any two points of an extended non-Euclidean space, as well as between two proper points when interpreting non-Euclidean space in a bundle of straight lines, is defined \textit{as the absolute value of the product of the angle between the corresponding straight lines of the bundle by the radius of curvature of the non-Euclidean space}.

Since the angle between an arbitrary line of Euclidean space and an isotropic line of this space is infinity, the distance from any point of non-Euclidean space to the point of the absolute is infinity as a result of which the points of the absolute of non-Euclidean space are also called \textit{infinitely distant points} of this space.

Points of non-Euclidean spaces $S_{p,q}$ both proper and infinitely distant and ideal, can be represented by arbitrary vectors of spaces $\R_{p+1,q}$ directed along straight bundles corresponding to these points. These vectors satisfy the condition
\begin{equation}
    (x,x)>0
\end{equation}
for proper points of the space $S_{p,q}$ of positive curvature and ideal points of the spaces $S_{p,q}$ of negative curvature. They satisfy the condition
\begin{equation}
    (x,x)<0
\end{equation}
for ideal points of spaces $S_{p,q}$ of positive curvature and proper points of spaces $S_{p,q}$ of negative curvature and satisfy the condition
\begin{eqnarray}\label{151}
    (x,x)=0
\end{eqnarray}
for points of absolutes of spaces $S_{p,q}$.

\section{Spinor Groups}
For an $n$-dimensional pseudo-Euclidean space $\R^{p,q}$ ($n=p+q$), the Lipschitz group
\[
\Lip_{p,q}=\left\{s\in\cl_{p,q}\;|\;\forall \bx\in\R^{p,q},\;
s\bx s^{-1}\in\R^{p,q}\right\}
\]
contains a spinor group
\[
\spin(p,q)=\left\{s\in\Lip^+_{p,q}\;|\;N(s)=\pm 1\right\},
\]
where $\Lip^+_{p,q}=\Lip_{p,q}\cap\cl^+_{p,q}$, $\cl^+_{p,q}$ is an even subalgebra of the Clifford algebra $\cl_{p,q}$ of the space $\R^{p,q}$. The group $\spin(p,q)$ contains the subgroup
\[
\spin_+(p,q)=\left\{s\in\spin(p,q)\;|\;N(s)=1\right\}.
\]

A twofold covering $\spin_+(p,q)$ of the rotation group
$\SO_0(p,q)$ of the pseudo-Euclidean space $\R^{p,q}$ is described
in terms of even subalgebra $\cl^+_{p,q}$. In its turn, the
subalgebra $\cl^+_{p,q}$ admits the following isomorphisms:
$\cl^+_{p,q}\simeq\cl_{q,p-1}$, $\cl^+_{p,q}\simeq\cl_{p,q-1}$. At this point, for the algebras of general type $\cl_{p,q}$ ($p\neq q$) and $\cl_{p,0}$ we have
\begin{equation}\label{Isom1}
\cl^+_{p,q}\simeq\cl_{q,p-1}.
\end{equation}
The isomorphism
\begin{equation}\label{Isom2}
\cl^+_{p,q}\simeq\cl_{p,q-1}
\end{equation}
takes place for the algebras of type $\cl_{0,q}$ and $\cl_{p,p}$ ($p=q$).
In dependence on dimension of the associated space $\R^{p,q}$, the
subalgebras $\cl_{p,q-1}$ and $\cl_{q,p-1}$ have even dimensions
($p+q-1\equiv 0\pmod{2}$) or odd dimensions ($p+q-1\equiv
1\pmod{2}$).

When $p+q-1\equiv 1\pmod{2}$ we have four types $q-p+1\equiv
1,3,5,7\pmod{8}$ ($p-q+1\equiv 1,3,5,7\pmod{8}$) of the subalgebras
$\cl_{q,p-1}$ ($\cl_{p,q-1}$). In this case a center $\bZ_{q,p-1}$
($\bZ_{p,q-1}$) of $\cl_{q,p-1}$ ($\cl_{p,q-1}$) consists of the
unit $\e_0$ and the volume element
$\omega=\e_1\e_2\cdots\e_{p+q-1}$. When $\omega^2=-1$ ($q-p+1\equiv
3,7\pmod{8}$ or $p-q+1\equiv 3,7\pmod{8}$) we arrive at the
isomorphisms
$\cl_{q,p-1}\simeq\C_{q+p-2}\simeq\C\left(2^{\frac{q+p-2}{2}}\right)$, $\cl_{0,p-1}\simeq\C_{p-2}\simeq\C\left(2^{\frac{p-2}{2}}\right)$ and $\cl_{0,q-1}\simeq\C_{q-2}\simeq\C\left(2^{\frac{q-2}{2}}\right)$.
The transition from $\cl_{q,p-1}\rightarrow\C_{q+p-2}$, $\cl_{0,p-1}\rightarrow\C_{p-2}$ and $\cl_{0,q-1}\rightarrow\C_{q-2}$ can be represented as the transition from the real coordinates in
$\cl_{q,p-1}$, $\cl_{0,p-1}$ and $\cl_{0,q-1}$ to complex coordinates $a+\omega b$ in
$\C_{q+p-2}$, $\C_{p-2}$ and $\C_{q-2}$, where $\omega=\e_1\e_2\cdots\e_{p+q-1}\in\cl_{q,p-1}$, $\omega=\e_1\e_2\cdots\e_{p-1}\in\cl_{0,p-1}$ and $\omega=\e_1\e_2\cdots\e_{q-1}\in\cl_{0,q-1}$. On the other
hand, when $\omega^2=1$ ($q-p+1\equiv 1,5\pmod{8}$ or $p-q+1\equiv
1,5\pmod{8}$) we arrive at the isomorphisms
\begin{eqnarray}
\cl_{q,p-1}&\simeq&\cl_{p-1,q-1}\oplus\cl_{p-1,q-1}\simeq
\R\left(2^{\frac{p+q-2}{2}}\right)\oplus\R\left(2^{\frac{p+q-2}{2}}\right),\nonumber\\
\cl_{0,p-1}&\simeq&\cl_{0,p-2}\oplus\cl_{0,p-2}\simeq
\R\left(2^{\frac{p-2}{2}}\right)\oplus\R\left(2^{\frac{p-2}{2}}\right),\nonumber\\
\cl_{0,q-1}&\simeq&\cl_{0,q-2}\oplus\cl_{0,q-2}\simeq
\R\left(2^{\frac{q-2}{2}}\right)\oplus\R\left(2^{\frac{q-2}{2}}\right)\nonumber
\end{eqnarray}
if $q-p+1\equiv 1\pmod{8}$ (or $p-1\equiv 1\pmod{8}$ and $q-1\equiv 1\pmod{8}$). When $q-p+1\equiv 5\pmod{8}$ (or $p-1\equiv 5\pmod{8}$ and $q-1\equiv 5\pmod{8}$) we have
\begin{eqnarray}
\cl_{q,p-1}&\simeq&\cl_{p-1,q-1}\oplus\cl_{p-1,q-1}\simeq
\BH\left(2^{\frac{p+q-4}{2}}\right)\oplus\BH\left(2^{\frac{p+q-4}{2}}\right),\nonumber\\
\cl_{0,p-1}&\simeq&\cl_{0,p-2}\oplus\cl_{0,p-2}\simeq
\BH\left(2^{\frac{p-4}{2}}\right)\oplus\BH\left(2^{\frac{p-4}{2}}\right),\nonumber\\
\cl_{0,q-1}&\simeq&\cl_{0,q-2}\oplus\cl_{0,q-2}\simeq
\BH\left(2^{\frac{q-4}{2}}\right)\oplus\BH\left(2^{\frac{q-4}{2}}\right).\nonumber
\end{eqnarray}
The transitions $\cl_{q,p-1}\rightarrow{}^2\R\left(2^{\frac{p+q-2}{2}}\right)$, $\cl_{0,p-1}\rightarrow{}^2\R\left(2^{\frac{p-2}{2}}\right)$, $\cl_{0,q-1}\rightarrow{}^2\R\left(2^{\frac{q-2}{2}}\right)$ and $\cl_{q,p-1}\rightarrow{}^2\BH\left(2^{\frac{p+q-4}{2}}\right)$, $\cl_{0,p-1}\rightarrow{}^2\BH\left(2^{\frac{p-4}{2}}\right)$, $\cl_{0,q-1}\rightarrow{}^2\BH\left(2^{\frac{q-4}{2}}\right)$
can be represented as the transition from the real coordinates in
$\cl_{q,p-1}$, $\cl_{0,p-1}$, $\cl_{0,q-1}$ to double coordinates $a+\omega b$ in
${}^2\R\left(2^{\frac{p+q-2}{2}}\right)$, ${}^2\R\left(2^{\frac{p-2}{2}}\right)$, ${}^2\R\left(2^{\frac{q-2}{2}}\right)$ and
${}^2\BH\left(2^{\frac{p+q-4}{2}}\right)$, ${}^2\BH\left(2^{\frac{p-4}{2}}\right)$, ${}^2\BH\left(2^{\frac{q-4}{2}}\right)$.

Further, when $p+q-1\equiv 0\pmod{2}$ we have four types
$q-p+1\equiv 0,2,4,6\pmod{8}$ ($p-q+1\equiv 0,2,4,6\pmod{8}$) of the
subalgebras $\cl_{q,p-1}$ ($\cl_{p,q-1}$). Let $p+q-1\geq 4$ and let
$m=(p+q-1)/4$ be an integer number (this number is always integer,
since $p+q-1\equiv 0\pmod{2}$). Then, at $i\leq 2m$ we have
\begin{eqnarray}
\e_{12\ldots 2m 2m+k}\e_{i}&=&(-1)^{2m+1-i}\sigma(i-l)
\e_{12\ldots i-1 i+1\ldots 2m 2m+k},\nonumber \\
\e_{i}\e_{12\ldots 2m 2m+k}&=&(-1)^{i-1}\sigma(i-l) \e_{12\ldots i-1
i+1\ldots 2m 2m+k}\nonumber
\end{eqnarray}
and, therefore, the condition of commutativity of elements
$\e_{12\ldots 2m 2m+k}$ and $\e_{i}$ is $2m+1-i\equiv i-1\pmod{2}$.
Thus, the elements $\e_{12\ldots 2m 2m+1}$ and $\e_{12\ldots 2m
2m+2}$ commute with all the elements $\e_{i}$ whose indices do not
exceed $2m$. Therefore, a transition from the algebra $\cl_{q,p-1}$
to algebras $\cl_{q+2,p-1}$, $\cl_{q,p+1}$ and
$\cl_{q+1,p}$ can be represented as a transition from the real
coordinates in $\cl_{q,p-1}$ to coordinates of the
form $a+b\phi+c\psi+d\phi\psi$, where $\phi$ and $\psi$ are
additional basis elements $\e_{12\ldots 2m 2m+1}$ and $\e_{12\ldots
2m 2m+2}$. The elements $\e_{i_{1}i_{2}\ldots i_{k}}\phi$ contain
the index $2m+1$ and do not contain the index $2m+2$. In its turn,
the elements $\e_{i_{1}i_{2}\ldots i_{k}}\psi$ contain the index
$2m+2$ and do not contain the index $2m+1$. Correspondingly, the
elements $\e_{i_{1}i_{2}\ldots i_{k}} \phi\psi$ contain both indices
$2m+1$ and $2m+2$. Analogously, we have a transition from the algebra $\cl_{0,p-1}$ to $\cl_{2,p-1}$, $\cl_{0,p+1}$ and $\cl_{1,p}$, from the algebra $\cl_{0,q-1}$ to $\cl_{2,q-1}$, $\cl_{0,q+1}$ and $\cl_{1,q}$, and from the algebra $\cl_{p,p-1}$ to $\cl_{p+2,p-1}$, $\cl_{p,p+1}$, $\cl_{p+1,p}$. Hence it immediately follows that general
elements of these algebras can be represented as
\begin{equation}
\cA_{\cl_{q+2,p-1}}=\cl^0_{q,p-1}\e_0+\cl^1_{q,p-1}\phi+\cl^2_{q,p-1}\psi+
\cl^3_{q,p-1}\phi\psi,\label{El1}
\end{equation}
\begin{equation}
\cA_{\cl_{q,p+1}}=\cl^0_{q,p-1}\e_0+\cl^1_{q,p-1}\phi+\cl^2_{q,p-1}\psi+
\cl^3_{q,p-1}\phi\psi,\label{El2}
\end{equation}
\begin{equation}
\cA_{\cl_{q+1,p}}=\cl^0_{q,p-1}\e_0+\cl^1_{q,p-1}\phi+\cl^2_{q,p-1}\psi+
\cl^3_{q,p-1}\phi\psi,\label{El3}
\end{equation}
\begin{equation}
\cA_{\cl_{2,p-1}}=\cl^0_{0,p-1}\e_0+\cl^1_{0,p-1}\phi+\cl^2_{0,p-1}\psi+
\cl^3_{0,p-1}\phi\psi,\label{El4}
\end{equation}
\begin{equation}
\cA_{\cl_{0,p+1}}=\cl^0_{0,p-1}\e_0+\cl^1_{0,p-1}\phi+\cl^2_{0,p-1}\psi+
\cl^3_{0,p-1}\phi\psi,\label{El5}
\end{equation}
\begin{equation}
\cA_{\cl_{1,p}}=\cl^0_{0,p-1}\e_0+\cl^1_{0,p-1}\phi+\cl^2_{0,p-1}\psi+
\cl^3_{0,p-1}\phi\psi,\label{El6}
\end{equation}
\begin{equation}
\cA_{\cl_{2,q-1}}=\cl^0_{0,q-1}\e_0+\cl^1_{0,q-1}\phi+\cl^2_{0,q-1}\psi+
\cl^3_{0,q-1}\phi\psi,\label{El7}
\end{equation}
\begin{equation}
\cA_{\cl_{0,q+1}}=\cl^0_{0,q-1}\e_0+\cl^1_{0,q-1}\phi+\cl^2_{0,q-1}\psi+
\cl^3_{0,q-1}\phi\psi,\label{El8}
\end{equation}
\begin{equation}
\cA_{\cl_{1,q}}=\cl^0_{0,q-1}\e_0+\cl^1_{0,q-1}\phi+\cl^2_{0,q-1}\psi+
\cl^3_{0,q-1}\phi\psi,\label{El9}
\end{equation}
\begin{equation}
\cA_{\cl_{p+2,p-1}}=\cl^0_{p,p-1}\e_0+\cl^1_{p,p-1}\phi+\cl^2_{p,p-1}\psi+
\cl^3_{p,p-1}\phi\psi,\label{El10}
\end{equation}
\begin{equation}
\cA_{\cl_{p,p+1}}=\cl^0_{p,p-1}\e_0+\cl^1_{p,p-1}\phi+\cl^2_{p,p-1}\psi+
\cl^3_{p,p-1}\phi\psi,\label{El11}
\end{equation}
\begin{equation}
\cA_{\cl_{p+1,p}}=\cl^0_{p,p-1}\e_0+\cl^1_{p,p-1}\phi+\cl^2_{p,p-1}\psi+
\cl^3_{p,p-1}\phi\psi,\label{El12}
\end{equation}
\begin{sloppypar}\noindent
where $\cl^i_{q,p-1}$, $\cl^i_{0,p-1}$, $\cl^i_{0,q-1}$, $\cl^i_{p,p-1}$, are the algebras with a
general element ${\cal A}=\sum^{2m}_{k=0}a^{i_{1}i_{2}\ldots i_{k}}
\e_{i_{1}i_{2}\ldots i_{k}}$. When the elements $\phi=\e_{12\ldots
2m 2m+1}$ and $\psi=\e_{12\ldots 2m 2m+2}$ satisfy the condition
$\phi^{2}=\psi^{2}=-1$ we see that the basis
$\{\e_{0},\,\phi,\,\psi,\, \phi\psi\}$ is isomorphic to a basis of
the quaternion algebra $\cl_{0,2}$, therefore, the elements
(\ref{El2}), (\ref{El5}), (\ref{El8}) and (\ref{El11}) are general elements of quaternionic
algebras. In turn, when the elements $\phi$ and $\psi$ satisfy
the condition $\phi^{2}=-\psi^{2}=1$, the basis
$\{\e_{0},\,\phi,\,\psi,\, \phi\psi\}$ is isomorphic to a basis of
the anti-quaternion algebra $\cl_{1,1}$, and the elements
(\ref{El3}), (\ref{El6}), (\ref{El9}) and (\ref{El12}) are general elements of
anti-quaternionic algebras. Further, when $\phi^{2}=\psi^{2}=1$ we
have a basis of the pseudo-quaternion algebra $\cl_{2,0}$, and the
elements (\ref{El1}), (\ref{El4}), (\ref{El7}) and (\ref{El10}) are general elements of
pseudo-quaternionic algebras.\end{sloppypar}

Let us define matrix representations of the quaternion units $\phi$
and $\psi$ as follows:
\[
\phi\longmapsto\begin{bmatrix} 0 & -1\\ 1 & 0\end{bmatrix},\quad
\psi\longmapsto\begin{bmatrix} 0 & i\\ i & 0\end{bmatrix}.
\]
Using these representations and (\ref{El2}), we obtain
\[
\cl_{p,q+1}\simeq\Mat_2(\cl_{p,q-1})=\begin{bmatrix}
\cl^0_{p,q-1}-i\cl^3_{p,q-1} & -\cl^1_{p,q-1}+i\cl^2_{p,q-1}\\
\cl^1_{p,q-1}+i\cl^2_{p,q-1} & \cl^0_{p,q-1}+i\cl^3_{p,q-1}
\end{bmatrix}.
\]
The analogous expression takes place for the algebra $\cl_{q,p+1}$
with the element (\ref{El5}) and so on.

On the other hand, the algebras $\cl_{p,q-1}$ ($\cl_{q,p-1}$) with
$p+q-1\equiv 0\pmod{2}$ admit factorizations with respect to the
subalgebras $\cl_{0,2}$, $\cl_{2,0}$ and $\cl_{1,1}$. In accordance
with \cite[Prop.~3.16]{Kar79}, when $p+q-1\equiv 0\pmod{2}$ and
$\omega^2=\e^2_{12\ldots p+q-1}=1$, then $\cl_{p,q-1}$
($\cl_{q,p-1}$) is called {\it positive} ($\cl_{p,q-1}>0$ and
$\cl_{q,p-1}>0$ at $p-q+1\equiv 0,4\pmod{8}$ and $q-p+1\equiv
0,4\pmod{8}$) and correspondingly {\it negative} if $\omega^2=-1$
($\cl_{p,q-1}<0$ and $\cl_{q,p-1}<0$ at $p-q+1\equiv 2,6\pmod{8}$
and $q-p+1\equiv 2,6\pmod{8}$). Hence it follows that if
$\cl(V,Q)>0$, and $\dim V$ is even, then $\cl(V\oplus
V^\prime,Q\oplus Q^\prime)\simeq
\cl(V,Q)\otimes\cl(V^\prime,Q^\prime)$, and also if $\cl(V,Q)<0$,
and $\dim V$ is even, then $\cl(V\oplus V^\prime, Q\oplus
Q^\prime)\simeq \cl(V,Q)\otimes\cl(V^\prime,-Q^\prime)$, where $V$
is a vector space associated with $\cl_{p,q}$, $Q$ is a quadratic
form of $V$. Using the Karoubi Theorem \cite[Prop.~3.16]{Kar79}, we obtain for the algebra
$\cl_{p,q+1}$ the following factorization:
\begin{equation}\label{Ten}
\cl_{p,q+1}\simeq\underbrace{\cl_{s_i,t_j}\otimes\cl_{s_i,t_j}\otimes\cdots
\otimes\cl_{s_i,t_j}}_{\frac{p+q+1}{2}\;\text{times}}
\end{equation}
where $s_i,t_j\in\{0,1,2\}$. The analogous factorizations take place
for the algebras $\cl_{p+2,q-1}$, $\cl_{p+1,q}$, $\cl_{q+2,p-1}$,
$\cl_{q,p+1}$ and $\cl_{q+1,p}$.

As is known, Lipschitz showed that rotations of the pseudo-Euclidean
spaces $\R^{p,q}$ are defined by spinor groups \cite{Lips}
\[
\spin(p,q)=\left\{s\in\Lip^+_{p,q}\;|\;N(s)=\pm 1\right\},
\]
where $\Lip^+_{p,q}=\Lip_{p,q}\cap\cl^+_{p,q}$ is the special
Lipschitz group, $s\in\cl^+_{p,q}$ is an even invertible element of
$\cl_{p,q}$. More precisely, the element $s$ is the linear
combination of basis elements with even number, that is,
\[
s=\sum_ka^{i_1\ldots i_{2k}}\e_{i_1\ldots i_{2k}}.
\]
The condition $N(s)=\pm 1$ means that
\begin{equation}\label{5.99}
\sum_k\sigma(i_1)\cdots\sigma(i_{2k})\left(a^{i_1\ldots
i_{2k}}\right)^2=\pm 1.
\end{equation}
On the other hand, from the fact that the elements
$s\in\Lip^+_{p,q}$ are even products of elements $\nu=\sum\nu^i\e_i$
it is easy to obtain that coordinates of $s\in\spin(p,q)$ are
related by the following conditions:
\begin{equation}\label{5.100}
a^{i_1i_2\ldots i_{2k}}(a)^{k-1}=(2k-1)!!a^{[i_1i_2}a^{i_2i_3}\cdots
a^{i_{2k-1}i_{2k}]}.
\end{equation}
The conditions (\ref{5.100}) can be rewritten in the form
\begin{equation}\label{5.101}
\left.\begin{array}{rcl}
aa^{i_1i_2i_3i_4}&=&3!!a^{[i_1i_2}a^{i_3i_4]},\\
aa^{i_1i_2i_3i_4i_5i_6}&=&5!!a^{[i_1i_2}a^{i_3i_4i_5i_6]},\\
\hdotsfor[2]{3}\\
aa^{i_1i_2\ldots i_{2k}}&=&(2k-1)!!a^{[i_1i_2\ldots}a^{i_3i_4\ldots
i_{2k-1}i_{2k}]}.
\end{array}\right\}
\end{equation}
The conditions (\ref{5.99}) and (\ref{5.100}) mean that all the
coordinates of the elements $s$, that belong the spinor group
$\spin(p,q)$, are expressed via $n(n-1)/2$ coordinates $a^{ij}$,
where $n=p+q$. The number of coordinates $a^{ij}$ coincides with the
number of parameters of the rotation group of the space $\R^{p,q}$.
This fact shows that expressions (\ref{5.99}) and (\ref{5.100}) form
a full system of conditions that eliminate the spinor group
$\spin(p,q)$ from the algebra $\cl_{p,q}$.

For example, let us consider the algebra $\cl_{4,4}$ associated with the eight-dimensional pseudo-Euclidean space $\R^{4,4}$. The twofold covering $\spin_+(4,4)$ of the rotation group $\SO_0(4,4)$ of $\R^{4,4}$ is described in terms of the even subalgebra $\cl^+_{4,4}$ (see the application of the group $\SO_0(4,4)$ to the periodic table of chemical elements \cite{Var25}). The algebra $\cl_{2,4}$ has the type $p-q\equiv 0\pmod{8}$, therefore, from (\ref{Isom2}) we have $\cl^+_{4,4}\simeq\cl_{4,3}$, where $\cl_{4,3}$ is the algebra associated with the seven-dimensional space $\R^{4,3}$. In its turn, $\cl_{4,3}$ has the type $p-q\equiv 1\pmod{8}$ and, therefore, there is an isomorphism
\[
\cl_{4,3}\simeq\cl_{3,3}\oplus\cl_{3,3}\simeq\R(8)\oplus\R(8)={}^2\R(8).
\]
Further, in virtue of (\ref{Ten}) the algebra $\cl_{3,3}$ admits the following factorization: $\cl_{3,3}\simeq\cl_{2,0}\otimes\cl_{2,0}\otimes\cl_{1,1}\simeq\cl_{2,0}\otimes\cl_{3,1}$. Therefore,
\[
\cA_{\cl_{3,3}}=\cl^0_{3,1}\e_0+\cl^1_{3,1}\phi+\cl^2_{3,1}\psi+
\cl^3_{3,1}\phi\psi,
\]
where $\phi=\e_{12345}$, $\psi=\e_{12346}$ are pseudoquaternionic units, $\phi^2=\psi^2=1$. Taking into account
\[
\phi\longmapsto\begin{bmatrix} 0 &  1\\ 1 & 0\end{bmatrix},\quad
\psi\longmapsto\begin{bmatrix} 1 & 0\\ 0 & -1\end{bmatrix},
\]
we obtain
\begin{equation}\label{8-Group}{\renewcommand{\arraystretch}{1.2}
\spin_+(4,4)=\left\{s\in{}^2\left.\begin{bmatrix} \cl^0_{3,1}+\cl^2_{3,1} &
\cl^1_{3,1}-\cl^3_{3,1}\\
\cl^1_{3,1}+\cl^3_{3,1} & \cl^0_{3,1}-\cl^2_{3,1}\end{bmatrix}\right|\;N(s)=1
\right\}.}
\end{equation}
Since $\cl_{3,1}\simeq\Mat_4(\R)$, then for the double matrix in (\ref{8-Group}) we have
\[
{}^2\begin{bmatrix} \cl^0_{3,1}+\cl^2_{3,1} &
\cl^1_{3,1}-\cl^3_{3,1}\\
\cl^1_{3,1}+\cl^3_{3,1} & \cl^0_{3,1}-\cl^2_{3,1}\end{bmatrix}\in\Mat_8(\R)\oplus\Mat_8(\R).
\]

\section{Interpretation of Spinor Representations}
Let's find a geometric interpretation of spinor representations of groups of motions of non-Euclidean spaces \cite{Rozen55}.

Consider the space $S_{p,q}$ ($p-q\equiv 0\pmod{8}$) with an absolute (see the condition (\ref{151}))
\begin{equation}\label{214}
    (x^0)^2-(x^1)^2+(x^2)^2+\ldots - (x^{p+1-1})^2+(x^{p+q})^2=0.
\end{equation}
Spinor representations of the motions of this space are carried out by the aggregates of algebra $\cl_{p+1,q}$, the single-index elements of which satisfy the conditions
\begin{equation}\label{215}
    \e^2_1=+1,\;\e^2_2=-1,\;\e^2_3=+1,\ldots,\e^2_{p+q-1}=+1,\;\e^2_{p+q}=-1.
\end{equation}
Let's represent these aggregates as elements of algebra $\cl_{p+2,q}$, which are linear combinations of basic elements with an even number of indices. Any such element can be written as
\begin{eqnarray}\label{216}
    \alpha+\alpha^1\e_{12\ldots p+q-2,p+q-1}+\alpha^2\e_{12\ldots p+q-2,p+q}+\alpha^{12}\e_{p+q-1,p+q},
\end{eqnarray}
where $\alpha$, $\alpha^1$, $\alpha^2$, $\alpha^{12}$ are aggregates of algebra $\cl_{p,q-1}$ represented by elements of algebra $\cl_{p+1,q-1}$.

Since the elements $1$, $\e_{12\ldots p+q-2,p+q-1}$, $\e_{12\ldots p+q-2,p+q}$, $\e_{p+q-1,p+q}$ form the basis of an algebra isomorphic to the algebra of antiquaternions, with
\[
\e^2_{12\ldots p+q-2,p+q-1}=+1,\quad \e^2_{12\ldots p+q-2,p+q}=-1,
\]
\[
\e_{p+q-1,p+q}=-\e_{12\ldots p+q-2,p+q-1}\e_{12\ldots p+q-2,p+q},
\]
the aggregates of algebra $\cl_{p+1,q}$ can be represented as
\begin{equation}\label{217}
    \begin{bmatrix}
        A & B\\
        C & D
    \end{bmatrix}=\begin{bmatrix}
        \alpha-\alpha^{12} & \alpha^1+\alpha^2\\
        \alpha^1-\alpha^2 & \alpha+\alpha^{12}
    \end{bmatrix}.
\end{equation}
Representing in the same way the elements $\alpha$, $\alpha^1$, $\alpha^2$, $\alpha^{12}$ of algebra $\cl_{p,q-1}$ in the form of second-order matrices, the elements of which are elements of algebra $\cl_{p-1,q-2}$, and continuing this process until we reach the real matrix, we get a real matrix of the $2^{p+q}$-th order representing an arbitrary element of algebra $\cl_{p+1,q}$.

The motions of the space $S_{p,q}$ can be represented by the transformation
\begin{equation}\label{218}
    \mathrm{H}=\mathrm{A}\Xi\mathrm{A}^{-1},
\end{equation}
where $\mathrm{A}$ is an element of the spinor group, and $\Xi$ and $\mathrm{H}$ are linear combinations of basic elements with $p+q$ indices:
\begin{equation}\label{219}
    \left.\begin{array}{ccl}
       \Xi & =  & \sum_ix^i\e_{12\ldots i-1 i+1\ldots p+q}, \\
        \mathrm{H} & = & \sum_iy^i\e_{12\ldots i-1 i+1\ldots p+q}\quad (i=0,1,\ldots,p+q).
    \end{array}\right\}
\end{equation}
Due to the fact that
\begin{equation}\label{220}
    \left.\begin{array}{rcl}
        \e^2_{12\ldots i-1 i+1\ldots p+q} &=& (-1)^i \\
         \e_{12\ldots i-1 i+1\ldots p+q}\e_{12\ldots j-1 j+1\ldots p+q}&=& -e_{12\ldots j-1 j+1\ldots p+q}\e_{12\ldots i-1 i+1\ldots p+q},
    \end{array}\right\}
\end{equation}
the square of the element $\Xi$ is equal to
\begin{equation}\label{221}
    \Xi^2= (x^0)^2-(x^1)^2+(x^2)^2+\ldots - (x^{p+1-1})^2+(x^{p+q})^2=0.
\end{equation}

The aggregate (\ref{217}) can have the form (\ref{219}) only if $\alpha^1$ and $\alpha^2$ are real numbers, and $\alpha^{12}$ can be written as
\begin{equation}\label{222}
    \xi=\sum_ix^i\e_{12\ldots i-1 i+1\ldots p+q-2}\quad(i=0,1,\ldots, p+q-2).
\end{equation}
Therefore, the matrix (\ref{217}) representing the aggregate, which we will also denote by the letter $\Xi$, has the form
\begin{equation}\label{223}
    \Xi=\begin{bmatrix}
        -\xi & x^{p+q-1}+x^{p+q}\\
        -x^{p+q-1}+x^{p+q} & \xi
    \end{bmatrix}.
\end{equation}

Representing the aggregate $\xi$ in the same way in the form of similar second-order matrices and continuing this process until we reach the real matrix, we get a second-order real matrix representing the aggregate $\Xi$, which we will also denote with the same letter $\Xi$. For $n=1,2,3$, the matrix $\Xi$ has the form, respectively
\begin{equation}\label{224}
    \Xi=\begin{bmatrix}
        -x^0 & x^1+x^2\\
        -x^1+x^2 & x^0
    \end{bmatrix},
\end{equation}
\begin{equation}\label{225}
    \Xi=\begin{bmatrix}
        x^0 & -x^1-x^2 & x^3+x^4 & 0\\
        x^1-x^2 & -x^0 & 0 & x^3+x^4\\
        -x^3+x^4 & 0 & -x^0 & x^1+x^2\\
        0 & -x^3+x^4 & -x^1+x^2 & x^0
    \end{bmatrix},
\end{equation}
\begin{equation}\label{226}
    \Xi=\begin{bmatrix}
        \scr-x^0 &\scr x^1+x^2 &\scr -x^3-x^4 &\scr 0 &\scr x^5+x^6 &\scr 0 &\scr 0 &\scr 0\\
       \scr -x^1+x^2 &\scr x^0 &\scr 0 &\scr -x^3-x^4 &\scr 0 &\scr x^5+x^6 &\scr 0 &\scr 0\\
       \scr x^0-x^4 &\scr 0 &\scr x^0 &\scr -x^1-x^2 &\scr 0 &\scr 0 &\scr x^5+x^6 &\scr 0\\
       \scr 0 &\scr x^3-x^4 &\scr x^1-x^2 &\scr -x^0 &\scr 0 &\scr 0 &\scr 0 &\scr x^5+x^6\\
       \scr -x^5+x^6 &\scr 0 &\scr 0 &\scr 0 &\scr x^0 &\scr -x^1-x^2 &\scr x^3+x^4 &\scr 0\\
       \scr 0 &\scr -x^5+x^6 &\scr 0 &\scr 0 &\scr x^1-x^2 &\scr -x^0 &\scr 0 &\scr x^3+x^4\\
       \scr 0 &\scr0 &\scr -x^5+x^6 &\scr 0 &\scr -x^3+x^4 &\scr 0 &\scr -x^0 &\scr x^1+x^2\\
       \scr 0 &\scr 0 &\scr 0 &\scr -x^5+x^6 &\scr 0 &\scr -x^3-x^4 &\scr -x^1+x^2 &\scr x^0
    \end{bmatrix}.
\end{equation}

Each matrix $\Xi$ corresponds to a point in space $S_{p,q}$ with coordinates $x^i$. Consider the matrices $\Xi$ corresponding to the points of the absolute of the space $S_{p,q}$. Since the coordinates of such a point satisfy condition (\ref{214}), the corresponding matrix $\Xi$ satisfies condition $\Xi^2=0$, which implies that its determinant is also $0$.Therefore, there are nonzero real spinors of the space $S_{p,q}$ that transform to zero with a linear transformation corresponding to this matrix, i.e. vectors $a$ for which
\begin{equation}\label{227}
    b=\Xi a=0.
\end{equation}
If we denote the coordinates of the vectors $a$ and $b$ by
\begin{equation}\label{228}
    \left.\begin{array}{cc}
         & a,\; a^1,\; a^2,\; a^{12},\; a^3,\; a^{13},\; a^{23},\; a^{123},\; \ldots,\; a^{12\ldots n},  \\
         & b,\; b^1,\; b^2,\; b^{12},\; b^3,\; b^{13},\; b^{23},\; b^{123},\; \ldots,\; b^{12\ldots n},
    \end{array}\right\}
\end{equation}
respectively, condition (\ref{227}) can be written in coordinates in the form of $2^n$ linear equations ($n=p+q$), which, for $n=1,2,3$, respectively, have the form
\begin{equation}\label{229}
    \left.\begin{array}{rcl}
       b  & = & -x^0a+(x^1+x^2)a^1=0,  \\
       b^1  & = & -(x^1-x^2)a+x^0a^1=0,
    \end{array}\right\}
\end{equation}
\begin{equation}\label{230}
    \left.\begin{array}{rcl}
        b & = & x^0a-(x^1x^2)a^1+(x^3+x^4)a^2=0, \\
        b^1 & = & (x^1-x^2)a-x^0a^1+(x^3+x^4)a^{12}=0,\\
        b^2 & = & -(x^3-x^4)a-x^0a^2+(x^1+x^2)aa^{12}=0,\\
        b^{12} & = & -(x^3-x^4)a^1-(x^1-x^2)a^2+x^0a^{12}=0,
    \end{array}\right\}
\end{equation}
\begin{equation}\label{231}
    \left.\begin{array}{rcl}
        b & = & -x^0a+(x^1+x^2)a^1-(x^3+x^4)a^2+(x^5+x^6)a^3=0,  \\
        b^1 & = & -(x^1-x^2)a+x^0a^1-(x^3+x^4)a^{12}+(x^5+x^6)a^{13}=0,\\
        b^2 & = & (x^3-x^4)a+x^0a^2-(x^1+x^2)a^{12}+(x^5+x^6)a^{23}=0,\\
        b^{12} & = & (x^3-x^4)a^1+(x^1-x^2)a^2-x^0a^{12}+(x^5+x^6)a^{123}=0,\\
        b^3 & = & -(x^5-x^6)a+x^0a^3-(x^1+x^2)a^{13}+(x^3+x^4)a^{23}=0,\\
        b^{13} & = & -(x^5-x^6)a^1+(x^1=x^2)a^3-x^0a^{13}+(x^3+x^4)a^{123}=0,\\
        b^{23} & = & -(x^5-x^6)a^2-(x^3-x^4)a^3-x^0a^{23}+(x^1+x^2)a^{123}=0,\\
        b^{123} & = & -(x^5-x^6)a^{12}-(x^3-x^4)a^{13}-(x^1-x^2)a^{23}+x^0a^{123}=0.
    \end{array}\right\}
\end{equation}

Let's find the conditions under which these systems of equations contain the minimal number of independent equations. The system of equations (\ref{229}) consists of two independent equations, the number of which cannot be reduced; these equations determine a point on the absolute of the plane $S_{1,1}$, which is a real curve.

The system of equations (\ref{230}) contains three independent equations, since equation $b^{12}=0$ is satisfied by virtue of the three previous ones due to the identical relation
\begin{equation}\label{232}
    ab^{12}=a^1b^2-a^2b^1+a^{12}b.
\end{equation}

The equations $b=0$, $b^1=0$, $b^2=0$ define a straight line consisting of the points of the absolute of the space $S_{2,2}$, on which the plane generators of the maximal dimension are straight lines.

The system of equations (\ref{231}) it can be reduced to four independent equations, since equations $b^{12}=0$, $b^{13}=0$, $b^{23}=0$ are satisfied by virtue of equations $b=0$, $b^1=0$, $b^2=0$, $b^3=0$ due to identically fulfilled relations (\ref{232}) and the relations obtained from it by replacing indices $12$ with indices $13$ and $23$. In addition, the equation $b^{123}=0$ is satisfied by virtue of the seven preceding ones due to the relation
\begin{equation}\label{233}
    ab^{123}=a^{12}b^3-a^{13}b^2+a^{23}b^1,
\end{equation}
which is fulfilled identically when the condition
\begin{equation}\label{234}
    aa^{123}=a^1a^{23}-a^2a^{13}+a^3a^{12}
\end{equation}
is fulfilled.

The equations $b=0$, $b^1=0$. $b^2=0$, $b^3=0$ define a two-dimensional plane consisting of the points of the absolute of the space $S_{3,3}$, on which the plane generators of the maximal dimension are two-dimensional planes.

Note that if we denote the coordinates $a^1$, $a^2$, $a^3$, $a^{123}$ by $a^{14}$, $a^{24}$, $a^{34}$, $a^{1234}$, respectively, condition (\ref{234}) coincide with condition
\[
aa^{1234}=a^{14}a^{23}-a^{24}a^{13}+a^{34}a^{12},
\]
which is a special case of condition (\ref{5.100}) for $n=4$.

Similarly, we will find that for any $n=p+q$ the system of equations (\ref{231}) it can be reduced to $n+1$ independent equations
\begin{equation}\label{235}
    b=(-1)^n\left[ax^0+\sum_i(-1)^ia^i(x^{2i-1}+x^{2i})\right]=0,
\end{equation}
\begin{equation}\label{236}
    b^i=(-1)^n\left[(-1)ia(x^{2i-1}-x^{2i})+a^ix^0
    +\sum_j(-1)^ja^{ij}(x^{2j-1}+x^{2j})\right]=0,
\end{equation}
where $a^{ij}=-a^{ji}$. In fact, the rest of the equations are satisfied by virtue of these equations under conditions, which, if we denote the coordinates $a^{i_1\ldots i_k}$ with an odd number of indices through by $a^{i_1\ldots i_k n+1}$, coincide with the conditions (\ref{5.100}).

The equations $b=0$, $b^1=0$, $\ldots$, $b^{p+q}=0$ define a $(p+q-1)$-dimensional plane consisting of the points of the absolute of the space $S_{p,q}$ ($p-q\equiv 0\pmod{8}$), on which the plane generators of maximal dimension are $(p+q-1)$-dimensional planes that make up one connected family.

If we normalize the coordinates $a^{i_1\ldots i_k}$ by the condition
\begin{equation}\label{237}
    \sum_{(i_1\ldots i_k)}(a^{i_1\ldots i_k})^2=1,
\end{equation}
then due to the fact that for these coordinates, if for an odd number of indices we denote them by $a^{i_1\ldots i_k n+1}$, the condition (\ref{5.100}) are fulfilled, these coordinates can be considered as the coordinates of the elements of the spinor group of the algebra $\cl_{p+1,q}$, and for each real spinor of the space $S_{p,q}$, the coordinates $a^{i_1\ldots i_k}$ of which are equal the coordinates of the elements of the spinor group of the algebra $\cl_{p+1,q}$, correspond to the plane generator of the maximal dimension of the absolute of the space $S_{p,q}$.

On the other hand, each such plane generator can be matched with a real spinor of the space $S_{p,q}$ of the specified type. Indeed, if we draw the hyperplane (\ref{235}) through our plane, then multiplying the equation of the absolute (\ref{214}), which can be rewritten as
\begin{equation}
    (x^0)^2-\sum_i\left[(x^{2i-1})^2-(x^{2i})^2\right]
    =(x^0)^2-\sum_i(x^{2i-1}-x^{2i})(x^{2i-1}+x^{2i})=0,\label{238}
\end{equation}
by $a$, and equation (\ref{235}) by $(-1)x^0$ and adding one with the other, we get the relation
\begin{equation}\label{239}
    \sum_i(x^{2i-1}+x^{2i})\left[a(x^{2i-1}-x^{2i})+(-1)^ia^ix^0\right]=0.
\end{equation}

This relation will be satisfied if we represent its left side as a quadratic form with a skew-symmetric matrix, i.e. we put
\begin{equation}\label{240}
    a(x^{2i-1}-x^{2i})+(-1)^ia^ix^0=\sum_j(-1)^{i+1}a^{ij}(x^{2j-1}-x^{2j}).
\end{equation}
Transferring both terms of equation (\ref{240}) to the left side, we see that these equations coincide with equations (\ref{236}). Therefore, equations (\ref{235}) and (\ref{236}) form a complete system of equations of our plane generator. Denoting the coordinates $a^i$ by $a^{i n+1}$, we can use formulas (\ref{5.99}) and (\ref{5.100}) to determine from the coordinates $a$, $a^i$ and $a^{ij}$ all the other coordinates $a^{i_1\ldots i_k}$ of the $\cl_{p+1,q}$ algebra aggregate belonging to the spinor group and compose the remaining equations (\ref{227}) of our plane generator, which are consequences of equations (\ref{235}) and (\ref{236}). The resulting aggregate is determined up to multiplication by $-1$. This reasoning is inapplicable only in the case when the plane generator lies in the hyperplane $x^0=0$, but any such plane can be obtained by a limiting transition from plane generators that do not lie in the hyperplane $x^0=0$, and an element of the spinor group of the algebra $\cl_{p+1,q}$ can be placed on such a plane, obtained by a limiting transition from element corresponding to planes that do not lying in the hyperplane $x^0=0$.

Thus, we have obtained that \textit{the spinor group of the algebra $\cl_{p+1,q}$ two-digitally represents a family of plane generators of maximal dimension of the absolute of the space $S_{p,q}$}, and it follows from the continuity of all functions encountered here that this correspondence is continuous.

Further, in the space of real spinors of the space $S_{p,q}$ there is a surface (the equations of which, if we denote the coordinates $a^{i_1\ldots i_k}$ with an odd number of indices through $a^{i_1\ldots i_kn+1}$ have the form (\ref{5.100}) and (\ref{237}), representing two-digitally the family of plane generators of the maximal dimension of the absolute of the space $S_{p,q}$. During the motions of the space $S_{p,q}$, when the matrices $\Xi$ undergo transformation (\ref{218}), the real spinors undergo transformation
\begin{equation}\label{241}
    {}^\prime a=\mathrm{A}a.
\end{equation}

Since, during the motions of the space $S_{p,q}$, the families of the plane generators of the absolute pass into themselves, the surface in the space of spinors depicting this family also passes into itself. The coordinates $a^{i_1\ldots i_k}$ of the spinors, which are the radius vectors of the points of this surface, can be considered as the coordinates of the corresponding plane generators of the absolute; we see that during the motions of the space $S_{p,q}$, these coordinates undergo transformation (\ref{241}), i.e. they are transformed linearly; these linear transformations form the spinor representation of the motion group of the space $S_{p,q}$.

\end{document}